\documentclass[journal=langd5,manuscript=article,]{achemso}

\usepackage[version=3]{mhchem} 
\usepackage{subfigure} 
\usepackage{float}
\usepackage[dvipsnames,usenames]{color}
\usepackage{booktabs}
\usepackage{soul}
\usepackage{xcolor}
\usepackage{amssymb}
\usepackage{setspace}
\title{Temperature-sensitive Soft Microgels at Interfaces: Air-Water versus Oil-Water}

\author{Steffen~Bochenek}
\affiliation{Institute of Physical Chemistry, RWTH Aachen University, Landoltweg 2, 52056 Aachen, Germany}
\author{Andrea~Scotti}
\affiliation{Institute of Physical Chemistry, RWTH Aachen University, Landoltweg 2, 52056 Aachen, Germany}
\author{Walter~Richtering}
\email{richtering@rwth-aachen.de}
\affiliation{Institute of Physical Chemistry, RWTH Aachen University, Landoltweg 2, 52056 Aachen, Germany}
\date{\today}

\begin{document}
\maketitle

\begin{singlespacing}
\begin{abstract}

The formation of smart emulsions or foams whose stability can be controlled on-demand by switching external parameters is of great interest for basic research and applications. An emerging group of smart stabilizers are microgels, which are nano- and micro-sized, three-dimensional polymer networks that are swollen by a good solvent. In the last decades, the influence of various external stimuli on the two-dimensional phase behavior of microgels at air- and oil-water interfaces has been studied. However, the impact of the top-phase itself has been barely considered. Here, we present data that directly address the influence of the top-phase on the microgel properties at interfaces. The dimensions of pNIPAM microgels are measured after deposition from two interfaces, \emph{i.e.}, air- and decane-water. While the total in-plane size of the microgel increases with increasing interfacial tension, the portions or fractions of the microgels situated in the aqueous phase are not affected. We correlate the area microgels occupy to the surface tensions of the interfaces, which allows to estimate an elastic modulus. In comparison to nanoindentation measurements, we observe a larger elastic modulus for the microgels. By combining compression, deposition, and visualization, we show that the two-dimensional phase behavior of the microgel monolayers is not altered, although the microgels have a larger total in-plane size at higher interfacial tension. 
\textbf{A peer reviewed and extended version of this preprint and the electronic supplementary information can be found under S.~Bochenek, A.~Scotti, W.~Richtering, \textit{Soft Matter}, 2020, DOI: 10.1039/d0sm01774d.}

\end{abstract} 

\section{Introduction}
Two-phase systems formed by two liquids or a gas and a liquid, \emph{e.g.}, emulsions and foams, are omnipresent in industrial processes, nature, and everyday life. To prevent phase separation emulsifiers or foamers are used, which adsorb to the interface and obstruct the interfaces from reaching molecular close contact. Typical emulsifiers (or foamers) are amphiphilic molecules, called surfactants, or hard colloids, \cite{Ram1904,Pick1907} such as silica or polystyrene particles. More recently, soft colloids, for example synthetic, non-amphiphilic 
macromolecules \cite{Oka96}, polymer particles \cite{Bin01,Ngai05}, or proteins \cite{Hai1981}, have been explored at liquid interfaces, too. 
One of the most successful soft objects at interfaces are nano- and micro-sized gels, so-called microgels. 

Microgels are three-dimensional cross-linked polymer networks, which are swollen by a good solvent. Based on the specific nature of the polymer, microgels are stimuli-responsive \cite{Pel00}; for example, microgels composed of the poly\emph{N}-iso\-propyl\-acryl\-amide (pNIPAM) have a volume phase transition temperature (VPTT) of $\approx$~32$^\circ$C at which their polymer network de-swells. \cite{Hes68} pNIPAM microgels have a dense-core-fuzzy-shell structure with a decreasing polymer content from the center to the periphery, \cite{Sti04FF} as a consequence of the faster consumption of the cross-linker \emph{N,N'}-\-methyl\-ene\-bis\-acryl\-am\-ide (BIS) during the synthesis. \cite{Kro17} Starting from this, incorporation of other polymers and more sophisticated synthesis routines lead to a broad variety of microgels systems with different architectures \cite{Kar06, Men09, Nic19} and responsivenesses to several environmental stimuli. \cite{Saw91,Tan82,Mer15}

Microgels are highly interfacial active and adsorb to liquid-gas \cite{Zha99}, liquid-liquid \cite{Ngai05}, and solid interfaces \cite{Carr10}. Similar to rigid colloids, they irreversibly adsorb to interfaces with desorption energies in the order of 10$^6$ $k_BT$. \cite{Mont14} But there are also some significant and advantageous distinctions from hard particles. First, their stimuli-responsiveness in bulk is inherited to their adsorbed state. Emulsions\cite{Bru09,Ngai05,Ngai06,Fuj05,Dan19} and foams\cite{Dup08,Hor18,Wud18,Fam15} formed by microgels can be broken at will by mild changes in the environmental conditions, for example by heating above the VPTT or changing the pH. Second, the solvent swollen, porous, and soft polymer network of the microgels can adapt and stretch upon adsorption. \cite{Bru09,Gei12,Sch11,Rey16,Pin14,Rey17,Rav17,Har19,Boc19,Fab19} Thus increasing the contact area with the interface to maximize the gain in interfacial energy. 

The combination of these characteristics leads to additional functions and rich phase behavior. Microgels have been employed to explore manifold interfacial phenomena and advanced applications. However, the coupling of particle elasticity, molecular architecture, and interfacial effects is also a source of complexity. Understanding the structure of single microgels, the two-dimensional macroscopic phase behavior of microgel monolayers, and the mechanism of demulsification or defoaming has been a hot topic in the last two decades. 

External stimuli, such as pH \cite{Gei12, Gei14, Rav17, Mas14}, salt concentration \cite{Mas14,Rav17,Sch20}, and temperature \cite{Boc19,Har19, Mal17}, but also intrinsic proprieties, \emph{e.g.}, different cross-linker contents \cite{Rey17,Rav17}, architectures \cite{Gei15,Scot19}, and incorporation of inorganic cores \cite{Rau17, Gei15}, have been investigated at interfaces in the literature. These studies were conducted on both air-water and oil-water interfaces. Although a comparison of the different studies indicates that there is no qualitative difference between the interfaces, \emph{i.e.}, air-water and alkane-water, there has been no quantitative and direct experimental investigation of the very same microgel system. 

The top-phase is a key factor to consider when the microscopic conformation of the microgels at interfaces is inspected. Similar to bulk, the swelling equilibrium of a neutral microgel at the interface is determined by the free energy of mixing and elasticity but also the free energy of the surface, \emph{i.e.},~surface tension. \cite{Pin14,Mont14} Likewise, the solubility of the polymer in the two immiscible phases plays a dominant role in determining the final shape of confined microgels. \cite{Sch11,Rum16} When polar organic liquids are used as top-phase, which are solvents for pNIPAM, the structural heterogeneity (core-corona structure) of adsorbed microgels is mitigated. \cite{Rum16} Additionally, one has to consider the low surface tension between water and polar organic liquids. \cite{SurfWaterOrganics} 

In this study, we directly address the in-plane dimensions of single microgels and the two-dimensional phase behavior of microgel monolayers as a function of the top-phase, i.e., at air-water and decane-water interfaces. For this, a well-characterized model system of pNIPAM microgels is used. \cite{Scot19, Boc19} Both air and decane are non-solvents for pNIPAM \cite{Gei14c, Sch11} and, therefore, changes to the adsorbed microgel properties can be predominately related to the difference in surface tension. Combined compression and deposition experiment were conducted, with subsequent visualization of the microgels. Isotherms show an earlier first increase of the surface pressure at lower concentrations. These experiments were also performed at temperatures above and below the VPTT of the microgels in bulk. Independently of the top-phase, microgels monolayers show the same temperature dependence. We measure the stretching of individual microgels, connect the results to the different surface tensions of the two interfaces, and estimate the elastic moduli of the microgels. The entirety of the data unambiguously shows no alteration in the two-dimensional phase behavior of the microgel monolayers, although the microgels have a larger total in-plane size at higher interfacial tension.

\section{Experimental}
\subsection{Materials}
For all interface experiments, ultrapure water (Astacus$^2$, membraPure GmbH, Germany) with a resistivity of 18,2 MOhm$\cdot$cm was used as a sub-phase. Decane (Merck KGaA, Germany) was used as oil-phase. The decane was filtered three times over basic aluminum oxide (90 standardized, Merck KGaA, Germany). The last filtration step was done just before the experiment. Pieces of an ultra-flat silicon wafer ($\approx$~1.1~cm x~6.0 cm, P\{100\}, NanoAndMore GmbH, Germany) were used for depositions. 
Aqueous solutions of microgel were mixed with propan-2-ol (Merck KGaA, Germany). Microgels were also dispersed in chloroform (Merck KGaA, Germany).

\subsection{Compression isotherms and Depositions}
Gradient Langmuir-Blodgett type depositions\cite{Rey16} were conducted according to Ref.~\cite{Boc19}. Compression of a microgel monolayer is combined with the deposition to a solid substrate. While the monolayer is compressed the substrate is lifted through the interface in between the barriers with an angle (25$^\circ$). This produces a gradient of packing density of the deposited film on the substrate. Subsequently, the depositions are imaged \emph{ex situ} with atomic force microscopy (AFM). Because the compression and deposition is carried out at the same time, and the angle of the substrate is know, the physical properties can be connected to the microstructure of the monolayer. 

Before each measurement, the Langmuir-Blodgett trough was carefully cleaned and a fresh interface (air-water or decane-water) was created. The through is temperature controlled by circulating thermostated water through its base using an external water bath. Microgels were spread from solutions with a concentration of 10 mg mL$^{-1}$, mixed with 20~vol$\%$ of propan-2-ol (decane-water measurements),  50~vol$\%$ of propan-2-ol (air-water measurements), or from a solution with a concentration of 1 mg mL$^{-1}$ in chloroform (air-water measurements). Aqueous solutions were mixed with propan-2-ol to facilitate spreading. After temperature equilibration, the microgels were added to the interface using a syringe. Gradient Langmuir-Blodgett type depositions and compression isotherms were conducted at $(20.0\pm0.5)~^\circ$C and $(40.0\pm0.5)~^\circ$C.

\subsection{Atomic force Microscopy}
The microgel monolayers were imaged using a Dimension Icon atomic force microscope with closed loop (Veeco Instruments Inc., USA, Software: Nanoscope 9.4, Bruker Co., USA) in tapping mode. As probe OTESPA tips with a resonance frequency of 300~kHz, a nominal spring constant of 26 N~m$^{-1}$ of the cantilever and a nominal tip radius of $<$ 7~nm (Opus by Micromasch, Germany) were used. The programmed move function was employed to capture images of 7.5~$\mu$m x 7.5~$\mu$m (512~pixels x 512~pixels) in a straight line along the gradient direction on the substrate every 250 or 500 $\mu$m. For each line, an image at the position of highest compression was taken to obtain a reference position. 

\subsection{Image Analysis}
For all images, the open-source analysis software \textit{Gwyddion} 2.54 was used to remove the tilt and fix zero height to the minimum z-value of the image. AFM images of the deposited dried microgel monolayers were analyzed with a custom-written Matlab script based on the image analysis routine of Ref.~\cite{Rey16}. The script was already used in Ref.~\cite{Boc19} and details can be found there. The number of microgels per area, the mean nearest neighbor distances, and the hexagonal ordering parameter were calculated for all images. 

The analysis of microgel monolayers was furthered by height profiles of the microgels in the dried state. Profiles were extracted through the apizes of the microgels and at different angles with respect to the fast scan direction. Multiple height profiles of microgels at the same interfacial concentration are summarized and aligned to the microgels' apizes to obtain averaged microgel profiles and to not bias the results. Representative averaged height profiles were calculated from at least 40 microgels ($\approx$ 120 profiles). The averaged profiles are presented with the standard deviations as the error. The height of the microgels at the apex, $H_{core}$, is determined from the maximum value on the y-axis using the Matlab function \emph{findpeaks}. $D_{core}$ is computed using the Matlab function \emph{knnsearch} with a threshold value of 1.5~nm above the background (zero height of the image). At the highest surface pressures, microgel cores start to overlap and multiple microgel centers are shown. Here, $D_{core}$ was calculated from the distance between the apices.

\section{Results and Discussion}

For the investigation of microgels at air-water and oil-water interfaces a suitable model system is required.  We decided to use pNIPAM-based microgels, which have already been study in previous publications. \cite{Scot19, Boc19} The synthesis protocol and characterization of the physical properties of these microgels in solution can be found in Ref.~\cite{Boc19}. Briefly, the pNIPAM microgels have been synthesized with approximately 5~mol\% cross-linker (BIS) content. They have a hydrodynamic radius, $R_h$, of (153~$\pm$~3)~nm at T = 20$^\circ$~C and of (84~$\pm$~2)~nm at T = 40$^\circ$~C; their VPTT is around 33$^\circ$~C. The microgels were synthesized with $\approx$ 2~mol\%  of \emph{N}-(3-aminopropyl) meth\-acryl\-amide hydro\-chloride (APMH). The primary amine moities of the incorporated APMH allow post-modification of the microgels, \emph{e.g.}, covalent labeling with fluorescent dyes. \cite{Lyo11} Consequently the microgels are positively charged in pure water. Small-angle neutron scattering (SANS) data show a decay of the polymer volume fraction from the center to the periphery of the microgels at 20$^\circ$~C and a box-like profile in the collapsed state (T = 40$^\circ$~C). The microgels have a narrow size distribution and a size polydispersity of (7 $\pm$~1)~\% as obtained from the fit to the SANS data. \cite{Boc19}

\subsection{Stretching of Individual Microgels and Estimation of the Elastic moduli.}

\begin{figure}[h!]
{\includegraphics[width=1\textwidth]{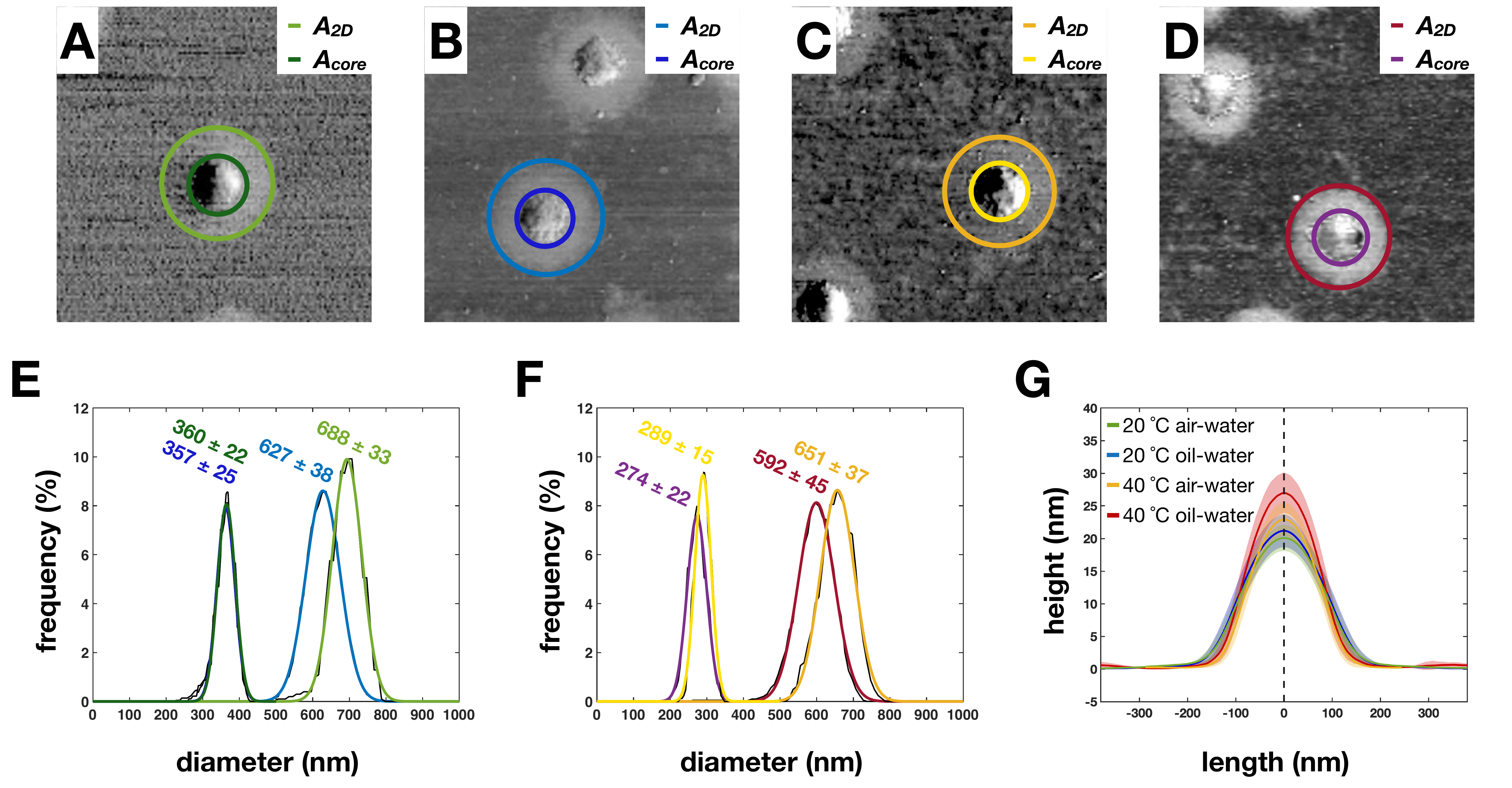}}
\centering
\caption{Interfacial dimensions of dried microgels at different temperatures and interfaces. (A-D) AFM phase images with circles indicating the area of the core and the whole microgel. (A) T = 20~$^\circ$C, (B) T = 20~$^\circ$C, (C) T = 40~$^\circ$C, and (D) T = 40~$^\circ$C. (E and F) Black lines are the size frequency distributions of pNIPAM microgels. Normal distributions are given in the corresponding colors of the circles in A-D, the labels give the mode and the variance. The diameter was calculated from the area with $D = 2\cdot\sqrt{A/\pi}$. (G) Averaged height profiles of microgels at T = 20. The standard deviations are given as shaded areas in the complimentary colors. For each state (A-D) at least 250 microgels were analyzed. 
}
\label{fig:fig_AFM_SingleMicrogels}
\end{figure}

We first evaluate the effect of the top-phase on the microgels by measuring the dimensions of individual, well-separated microgels after deposition with \emph{ex situ} AFM in the dried state. \cite{Fab19, Boc19}  The depositions were performed at interfacial concentrations far below the contact of the microgels. A clear contrast between the microgels and the silica wafer, but also between the core and corona can be seen in the phase contrast images of the AFM measurements. \cite{Har19,Boc19} Examples of the phase images are presented in Figure~\ref{fig:fig_AFM_SingleMicrogels}A-D. Circles representing the cores and the overall microgels are added for clarity. From the images we estimate the size distributions for the overall and core diameters (Figure~\ref{fig:fig_AFM_SingleMicrogels}E and F, black lines). The averaged overall and core diameters, $D_{2D}$ and $D_{core}$, respectively, are determined by fitting Gaussians to the data. The fits are plotted with the respective colors of Figure~\ref{fig:fig_AFM_SingleMicrogels}A-D. 

The corresponding AFM height images are used to extract topographic information of the microgels. We averaged the height profiles of multiple microgels at the same state (interface and temperature) to determine the height, $H_{core}$, and core diameter, $D_{core}$, of the dried microgels. 
The experiments were conducted with the very same pNIPAM microgels, Langmuir-Blodgett setup, and the same type of sub-phase and substrates. Thus, differences in the profiles 
must stem from conformational differences of the microgels at the air-water or oil-water interface, similar to the case of microgels deposited from bulk and air-water interfaces. \cite{Har19}  $D_{core}$ was determined twice, from respective height and phase images. Both methods yield the same $D_{core}$-values within the errors. The results from the phase images and height profiles are summarized in Table~\ref{tab:tab_NND}. 

Figure~\ref{fig:fig_AFM_SingleMicrogels} shows that at the same temperatures, $D_{2D}$ of the microgels increases from the decane-water to the air-water interface by $\approx$ 60~nm. At 20~$^\circ$C, $D_{2D}$ increases from 627~$\pm$~38 to 688~$\pm$~33~nm and at 40~$^\circ$C from 592~$\pm$~45 to 651~$\pm$~37~nm. In contrast, $D_{core}$ stays constant (within the errors) from oil-water to air-water interfaces below (357~$\pm$~25 to 360~$\pm$~22~nm) and above the VPTT (274~$\pm$~22 to 289~$\pm$~15~nm). A decrease of $H_{core}$ is observed for microgels below (from 21~$\pm$~2 to 20~$\pm$~2~nm) and above (from 27~$\pm$~3 to 23~$\pm$~2~nm) the VPTT from the decane- to the air-water interface (Tab.~\ref{tab:tab_NND}). 

\begin{figure}[h!]
{\includegraphics[width=\textwidth]{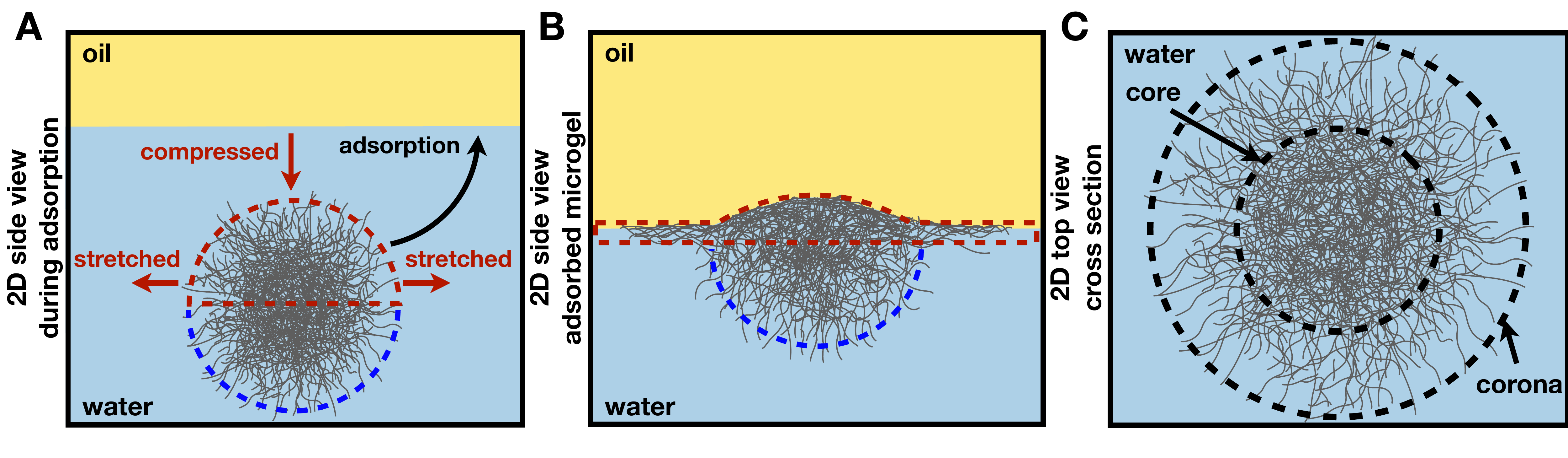}}
\centering
\caption{Sketch of the conformation of a pNIPAM microgel with inhomogeneous cross-linker distribution at the oil-water interface. (A) Side view during adsorption, (B) Side view after adsorption, and (C) 2D top view of a cross-section along the interfacial plane. For further explanation see text. 
}
\label{fig:fig_Sketch}
\end{figure}

To understand these changes in the dimensions ($D_{2D}$, $D_{core}$, and $H_{core}$) of the dried microgels we have to consider the adsorption process to and the structure of microgels at interfaces before deposition. When a microgel adsorbs to an interface the polymer network is able to adapt and deform to increase their contact area and maximize the gain in interfacial energy. Thereby, fractions of its polymer network are compressed and stretched (Fig.~\ref{fig:fig_Sketch}A and B, outlined in red) while other fractions do not experience significant deformation (Fig.~\ref{fig:fig_Sketch}A and B, outlined in blue). Due to this and to their inhomogeneous cross-linker distribution, microgels display a core-corona (or ``fried-egg'' like) structure at interfaces. \cite{Gei12} In direct contact and in close proximity to the interface the polymer network becomes denser and collapses (Fig.~\ref{fig:fig_Sketch}B, outlined in red). \cite{Fab19,Zie16,Kyr19,Son19} 
The fractions or portions of the microgels which are positioned further away from the interface (Fig.~\ref{fig:fig_Sketch}B, outlined in blue) keep similar properties to microgels in the bulk aqueous phase, \emph{e.g.}, they can still de-swell as a function of temperature. \cite{Mae18, Boc19} The sketch in Fig.~\ref{fig:fig_Sketch}B illustrates another crucial fact: The polymer layer in the interfacial plane consists of both dangling chains and cross-linked microgels' fractions. In other words, a thin cross-section of the microgel along interface (Fig~\ref{fig:fig_Sketch}C) displays a decrease of cross-linking and polymer distribution from the center to the periphery.

The larger $D_{2D}$ at air-water interfaces is not surprising as the microgels stretch further with increasing surface tension \cite{Rum16} from the decane-water interface (52.3~mN~m$^{-1}$ at 20~$^\circ$C) \cite{SurfWaterDecaneTemp} to the air-water interface (72.3~mN~m$^{-1}$ at 20~$^\circ$C) \cite{SurfWaterTemp}. In contrast, we observe that the size of the core ($D_{core}$) is independent of the top-phase. 
Considering the sketches in Fig.~\ref{fig:fig_Sketch} and the literature, \cite{Fab19, Gei15, Son19, MFS19, Har19} this might be explained by: (i) the fact that the polymer network in the center of the microgel gives more resistance to deformation due to the higher cross-linker content, and experiences less stress due to the higher polymer content (Fig.~\ref{fig:fig_Sketch}C). Consequently, the stretching of the core could be already ``saturated'' \cite{Fab19} or so small that it is not registered by our analysis.

(ii) Attenuation of the stretching for the microgels' portions that are in the aqueous phase due to the flexibility of the polymer network. In close contact with the interface, portions of the microgels are stretched due to the tensile stress (Fig.~\ref{fig:fig_Sketch}), but the stress can only be passed on through the cross-linking points to portions of the network further away from the interface. With increasing distance, the deformation is mitigated by the flexible polymer segments between the interconnection points. As a result, the portions situated in the aqueous phase are not (further) deformed.

(iii) Adsorption of more polymer sub-chains from the third dimension, \emph{i.e.},~out-of the interfacial plane, resulting in a stronger compression normal to the interface (Fig.~\ref{fig:fig_Sketch}A). In other words, instead of stretching the polymer network more in-plane, the higher surface tension is compensated by adsorbing more of the polymer network situated in the aqueous phase (Fig.~\ref{fig:fig_Sketch}B, outlined in blue) leading to a thinner microgel center. The height profiles in Figure~\ref{fig:fig_AFM_SingleMicrogels}G show a decrease in $H_{core}$ from the oil-water to the air-water interfaces, however, the changes are within the experimental uncertainties, and a clear statement cannot be made.    

Regardless of which explanation or combination of explanations apply, we clearly observe an increase in $D_{2D}$, while the lateral expansion of the core remains constant. A similar comparison by Camerin~\emph{et al.} between microgels at benzene- and hexane-water interfaces was conducted using both numerical and experimental results. \cite{Fab19} In their work they observe no significant variation of the in-plane size of the microgels. It is proposed that the stretching of the microgels is already ``saturated'' between the surface tension of benzene- (35.0~mN~m$^{-1}$ at 20~$^\circ$C) \cite{SurfWaterOrganics} and hexane-water (51.1~mN~m$^{-1}$ at 20~$^\circ$C) \cite{SurfWaterOrganics}, contradicting the results presented here. However, in the experiments, Camerin~\emph{et al.} used larger microgels. Recently, for the pH response of adsorbed microgels, the size of the system has been shown to play a pivotal role on the in-plane dimensions \cite{Sch20}. 
Consequently, further experiments have to be conducted, using different oils and microgels of different size and softness to clarify this discrepancy.

$D_{2D}$ as a function of the surface tension can be used to estimate the mean surface elasticity and its counterpart in bulk, the elastic modulus, of the whole microgel. Since in this experiment the spreading of individual microgels after adsorption is compared, we start with the equation of the hydrostatic compression in two dimensions, defined as \cite{Beh96}:
\begin{equation}
    K \equiv - \frac{\Delta p}{\Delta s/s}\,,
\end{equation}
\noindent where $K$ is the two-dimensional modulus of compression, $p$ is the two-dimensional analog of pressure (in force per unit length), and $s$ area of the interface. Hydrostatic compression states that the film is compressed uniformly in all directions. \cite{Beh96} For our case, we can rewrite this equation into:
\begin{equation}
    K \equiv \frac{\Delta \gamma}{\Delta A_{2D}/A_{2D}}\,,
     \label{eq:compression_moduli}
\end{equation}
\noindent where $\gamma$ is the surface tension and $A_{2D}$ is the total area calculated from $D_{2D}$ at the air- and decane-water interface. We omit the negative sign as the force is in the opposite direction, \emph{i.e.},~an expansion is performed. With Equation~\ref{eq:compression_moduli}, a two-dimensional modulus of compression of $K$~= (96~$\pm$~15)$\,\cdot\,10^{-3}$~N~m$^{-1}$ is calculated for the microgels at the air- and decane-water interface.
 
The compression modulus in two dimensions and the surface elasticity are connected by the following relation: $E^{(2D)} = 2K(1-\nu^{(2D)})$ \cite{Tho92}. Here, $\nu^{(2D)}$ is the two-dimensional Poisson's ratio and $E^{(2D)}$ is the surface elasticity or two-dimensional elastic modulus. The two-dimensional Poisson's ratio can be obtained from the three-dimensional one assuming 
plane stress and 
plane strain conditions. \cite{Tho92} As aforementioned, it is observed that (i) $D_{2D}$ changes with $\gamma$, but $D_{core}$ does not, and (ii) in literature it has been shown that the compressibility at larger microgel-to-microgel distances is independent of the polymer network portions in the aqueous phase. \cite{Boc19} Thus, we assume that for the stretching of the individual microgels only plane stress conditions are valid, where the polymer at the interface dominates the elasticity. This assumption is in agreement with recent literature \cite{Fab20} and, therefore, $\nu^{(2D)} = \nu^{(3D)}$ \cite{Tho92} is used. This relation leads to the equation:
\begin{equation}
    \frac{E^{(2D)}}{2-2\nu^{(2D)}} = K \equiv \frac{\Delta \gamma}{\Delta A_{2D}/A_{2D}}~.
     \label{eq:elasticity_moduli}
\end{equation}
\noindent In the literature, $\nu^{(3D)}$ values are reported in the range between $\approx$ 0.25 to 0.5 for pNIPAM gels. \cite{Sch10, Has09, Vou13, Boo17} For a perfectly incompressible isotropic material ($\nu^{(2D)}$ = $\nu^{(3D)}$ = 0.5), Equation~\ref{eq:elasticity_moduli} becomes the same as Hooke's law. Using the reported Poisson's ratios, we compute surface elasticities in the range of 96$\,\cdot\,$10$^{-3}$~N~m$^{-1}$ $<$ $E^{(2D)}$ $<$ 140$\,\cdot\,$10$^{-3}$~N~m$^{-1}$. For further calculations we use $E^{(2D)}$ of (115~$\pm$~18)$\,\cdot\,10^{-3}$~N~m$^{-1}$ calculated with $\nu^{(3D)}$ =~0.4 \cite{Vou13}.

In the literature, \cite{Sch10, Bac18,Tag08} typically the Young's Moduli of microgels are presented. For comparison, we convert the surface elasticity in 2D into $E^{(3D)}$. As we assumed that the stress normal to the interfacial plane is zero (plane stress), the relation between $E^{(2D)}$ and $E^{(3D)}$ is given by \cite{Fab20}: $E^{(3D)} = E^{(2D)}/d$, where $d$ is the thickness perpendicular to the interface. Although the maximal extension of the polymer network into the aqueous phase can be measured, for example, with ellipsometry \cite{Mae18, Boc19}, we follow the suggestion of Camerin \emph{et al.} \cite{Fab20} and consider $d$ to be significantly smaller, in the order of a few nm. This is also in good agreement with the assumption that the polymer at the interface dominates the elasticity and the discussion about the structure of adsorbed microgels (Fig.~\ref{fig:fig_Sketch}B). We calculate $E^{(3D)}$ using $d$~= 5\,$\cdot$\,10$^{-9}$~m according to the film thickness of linear pNIPAM polymer measured with ellipsometry and neutron reflectometry at the air-water interface, \cite{Oku14,Sai95,Ric00,Zie16} and obtain a Young's modulus of $E^{(3D)}$ = (23~$\pm$~5)~MPa.

Finally, we can compare the value for $E^{(3D)}$ to the E-moduli measured by AFM normal to the interface. Thereby, one has to consider that the adsorbed layer is de-swollen and collapsed, containing a low amount of solvent. In comparison to the values in the literature for E-moduli above the VPTT (100~kPa to 13~MPa), \cite{Sch10, Bac18,Tag08} our result is one to two orders of magnitude larger, showing that adsorbed microgels are stiffer within the interfacial plane. Recently, Camerin~\emph{et al.} addressed the microgel-to-microgel interactions at interfaces. \cite{Fab20} They calculate larger E-moduli for the lateral compression of adsorbed microgels compared to the bulk.
Their data depicted that the in-plane stiffening of the microgels is attributable to the presence of the interface and the resulting restrain of the confined polymer chains.

\subsection{Two-dimensional Phase Behavior of Microgel Monolayers.}

The two-dimensional phase behavior of microgel monolayers at air- and decane-water interfaces was captured by gradient Langmuir-Blodgett type depositions, which combine compression with simultaneous deposition and subsequently imaging in dried state. \cite{Rey16} The positions the images on the substrate are than encode with the barrier positions, and thus, to the surface pressure, $\Pi$. The images are analyzed with a custom-written Matlab script to determine the number concentration, $N_{area}$, the center-to-center distances, \emph{NND}, and hexagonal order parameter, $\Psi_6$ (see Ref.\cite{Boc19} for detailed description of the Matlab routine).

\begin{figure}[h!]
{\includegraphics[width=\textwidth]{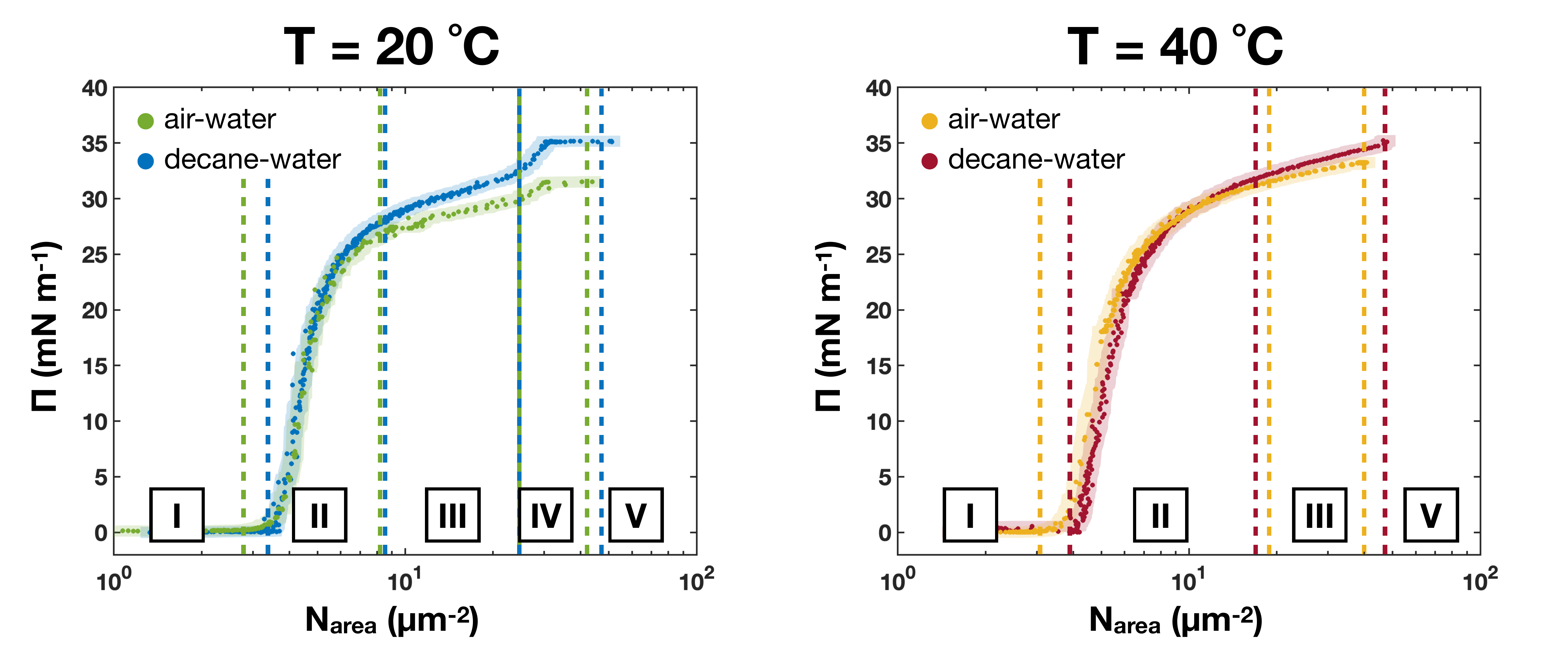}}
\centering
\caption{$\Pi$-$N_{area}$ compression isotherms of pNIPAM microgels at the air-water and decane-water interface. (left) Isotherms at T~=~20~$^\circ$C. (right) Isotherms at T~=~40~$^\circ$C. The errors are presented as shaded areas. The different regimes visible in the isotherms are labeled in roman numbers. Dashed lines in the respective colors indicate the transitions. $\Pi$-$area/mass$ compression isotherms are presented in the Supporting Information (Fig.~S1).
}
\label{fig:fig_Chapter5_Isotherm_NA}
\end{figure}

\begin{figure}[h!]
{\includegraphics[width=\textwidth]{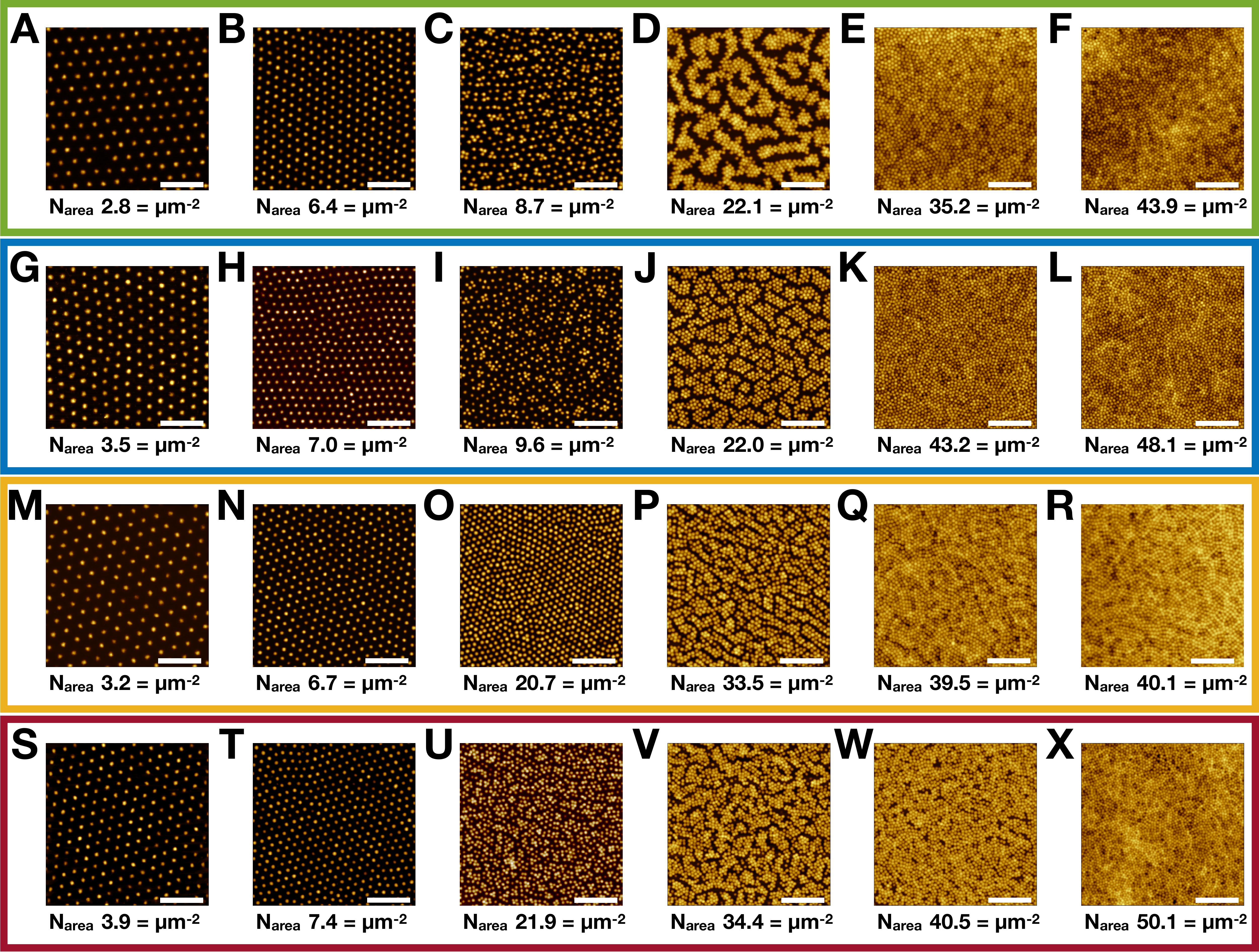}}
\centering
\caption{Atomic force micrographs of the microgel monolayers deposited at T = 20~$^\circ$C from the air-water interface (A-F, green box) and the decane-water interface (G-J, blue box), and deposited at 40~$^\circ$C from the air-water interface (M-R, golden box) and decane-water interface (S-X, red box) at different compressions. The scale bars are 2 $\mu$m.
}
\label{fig:fig_AFM_images}
\end{figure}

$\Pi$-$N_{area}$ compression isotherms of the pNIPAM microgels measured at both interfaces and at temperatures below and above their VPTT in solutions are presented in Figure~\ref{fig:fig_Chapter5_Isotherm_NA}. 
All compression isotherms show distinct regimes, in agreement with the literature \cite{Pin14,Rey16,Boc19,Rey17}. The transitions between the regimes are indicated by dashed lines in the corresponding colors of the isotherms. Below the VPTT (Fig.~\ref{fig:fig_Chapter5_Isotherm_NA} left) microgels display five regimes labeled in roman numbers from I to V, namely a diluted state (I), the corona-to-corona contact (II), the isostructural phase transition region (III), the core-to-core contact (IV), and the failure of the monolayer (V). The microgel monolayers at temperature above the VPTT (Fig.~\ref{fig:fig_Chapter5_Isotherm_NA} right) display only a single increase and regime IV, where microgels in core-to-core contact are compressed, is missing. A detailed discussion of the effect of temperature on microgel monolayers can be found in our previous work. \cite{Boc19} The regimes of the monolayers are further highlighted by AFM images below (Fig.~\ref{fig:fig_AFM_images}A-F, M-R) and above (Fig.~\ref{fig:fig_AFM_images}G-L, S-X) the VPTT. In general, the monolayers at the air-water interface display the same two-dimensional phase behavior as their temperature equivalents at the decane-water interface.

However, at all temperatures, the microgels monolayers at air-water interfaces have a transition at lower $N_{area}$-values from separated (regime I) to closed monolayers (regime II). AFM images of the first observed closed layers are presented in the first column of Figure~\ref{fig:fig_AFM_images}. At 20~$^\circ$C, the closed film is formed at 2.8~$\mu$m$^{-2}$ and at 3.5~$\mu$m$^{-2}$ at the air- and decane-water interfaces, respectively. For the monolayers above the VPTT (Fig~\ref{fig:fig_AFM_images}G and S), the same difference of $N_{area}$ is observed between the two top-phases. The microgels at the air-water interfaces occupy a larger area compared to the decane-water interfaces, confirming the above presented size difference measured for the individual microgels.   

\begin{table}[h!]
\centering
 \caption{Interfacial dimensions of microgels at different compressions deposited from air- and decane-water interfaces at T = 20 and 40~$^\circ$C. Center-to-center distance for the first phase, \emph{NND}$_{1st}$, the second phase, \emph{NND}$_{2nd}$, the in-plane diameter, $D_{2D}$, the core diameter, $D_{core}$, from phase and height images, and height of the dried microgels, $H_{core}$.}
\resizebox{\textwidth}{!}{\begin{tabular}{ccccccccc}
    \toprule
     Interface & $\Pi$ &$N_{Area}$ & \emph{NND}$_{1st}$ & \emph{NND}$_{2nd}$ & $D_{2D}$ & $D_{core}$$^a$ & $D_{core}$$^b$ & $H_{core}$ \\
     - & (mN~m$^{-1}$) & ($\mu$m$^{-2}$) & (nm) & (nm)  & (nm) & (nm)  & (nm)  & (nm) \\
      \midrule
       \multicolumn{9}{c}{T = 20~$^\circ$C} \\
     \midrule
        & 0 &$<$~2.8 & - & - & (688 $\pm$~33) & (360 $\pm$~22) & (369 $\pm$~35) & (20 $\pm$~2) \\
        & 0.2-1 & 2.8 & (620 $\pm$~20) & - & - & - & (328 $\pm$~34) & (19 $\pm$~2) \\
        & 10 & 4.2 & (500 $\pm$~25) & - & - & - & (304 $\pm$~30) & (22 $\pm$~2) \\
        & 20 & 5.3 & (450 $\pm$~30) & - & - & - & (292 $\pm$~30) & (26 $\pm$~3) \\
     air-water & 26 & 7.6 & (390 $\pm$~40) & - & - & - & (286 $\pm$~27)& (31 $\pm$~3)  \\
        & 26.5 & 8.1 & (380 $\pm$~35) & (220 $\pm$~20) & - & - & - & -  \\
        & 30 & 23.2 & (335 $\pm$~55) & (185 $\pm$~20) & - & - & (163 $\pm$~43) & (33 $\pm$~5)  \\
        & 31.2 & 29.0 & - & (191 $\pm$~12) & - & - & - & -  \\
        & 32.2 & 43.9 & - & (160 $\pm$~12) & - & - & - & -  \\
    \midrule
        & 0 &$<$~3.5 & - & - & (627 $\pm$~38) & (357 $\pm$~25) & (358 $\pm$~25) & (21 $\pm$~2)  \\
        & 0.2-1 &3.5 & (575 $\pm$~30) & - & - & - & (363 $\pm$~36) & (20 $\pm$~2)  \\
        & 10 & 4.8 & (495 $\pm$~20) & - & - & - & (294 $\pm$~30) & (28 $\pm$~2) \\
        & 20 & 5.8 & (450 $\pm$~30) & - & - & - & (272 $\pm$~25) & (35 $\pm$~3) \\
     decane-water & 26 & 7.0 & (385 $\pm$~25) & - & - & - & (266 $\pm$~26) & (41 $\pm$~3) \\
        & 27.5 &8.3 & (365 $\pm$~35) & (220 $\pm$~15) & - & - & - & -  \\
        & 30 & 15.3 & (330 $\pm$~35) & (230 $\pm$~15) & - & - & (181 $\pm$~19) & (43 $\pm$~2) \\
        & 32.1 & 25.8 & - & (182 $\pm$~22) & - & - & - & - \\
        & 34.7 & 48.1 & - & (146 $\pm$~12) & - & -  & - & - \\
      \midrule
       \multicolumn{9}{c}{T = 40~$^\circ$C} \\
     \midrule
        & 0 &$<$~3.2 & - & - & (651 $\pm$~37)  & (289 $\pm$~15) & (277 $\pm$~23) & (23 $\pm$~2)  \\
         & 0.2-1 &3.2 & (615 $\pm$~20) & - & - & - & (289 $\pm$~31) & (22 $\pm$~2)  \\
         & 20 & 5.6 & (445 $\pm$~20) & - & - & - & (253 $\pm$~25) & (29 $\pm$~3) \\
         & 30 & 12.5 & (300 $\pm$~20) & - & - & - & (245 $\pm$~26) & (34 $\pm$~3)  \\
      air-water & 31.3 & 18.0 & (255 $\pm$~20) & (185 $\pm$~20) & - & - & - & -  \\
         & 32 & 23.0 & (250 $\pm$~35) & (190 $\pm$~20) & - & - & (249 $\pm$~23) & (35 $\pm$~3) \\
         & 32.2 & 24.5 & (235 $\pm$~35) & (185 $\pm$~20) & - & - & (167 $\pm$~23) & (35 $\pm$~3) \\
         & 33.2 & 39.5 & (185 $\pm$~15) & (159 $\pm$~15) & - & - & - & -  \\
         & 33.2 & 40.1 & - & (162 $\pm$~14) & - & - & - & - \\
    \midrule
         & 0 &$<$~3.9 & - & - & (592 $\pm$~45)  & (274 $\pm$~22) & (280 $\pm$~30) & (27 $\pm$~3)  \\
         & 0.2-1 &3.9 & (540 $\pm$~25) & - & - & - & (265 $\pm$~25) & (26 $\pm$~3)  \\
         & 20 & 5.2 & (470 $\pm$~25) & - & - & - & (245 $\pm$~24) & (36 $\pm$~4) \\
     decane-water  & 30 & 10.3 & (320 $\pm$~25) & - & - & - & (239 $\pm$~20) & (44 $\pm$~3) \\
         & 32 & 17.0 & (270 $\pm$~30) & (168 $\pm$~20) & - & - & (230 $\pm$~30) & (45 $\pm$~5)  \\
         & 33.2 & 23.7 & (250 $\pm$~35) & (158 $\pm$~20) & - & - & (156 $\pm$~30) & (49 $\pm$~3) \\
         & 34.2 & 40.5 & (187 $\pm$~15) & (152 $\pm$~15) & - & - & - & -  \\
         & 34.8 & 47.8 & - & (146 $\pm$~15) & - & - & - & -  \\
     \bottomrule
     \multicolumn{9}{l}{$^a$ values determined from phase images}\\
     \multicolumn{9}{l}{$^b$ values determined from height profiles}\\
     \label{tab:tab_NND}
\end{tabular}}
\end{table}

\begin{figure}[h!]
{\includegraphics[width=0.9\textwidth]{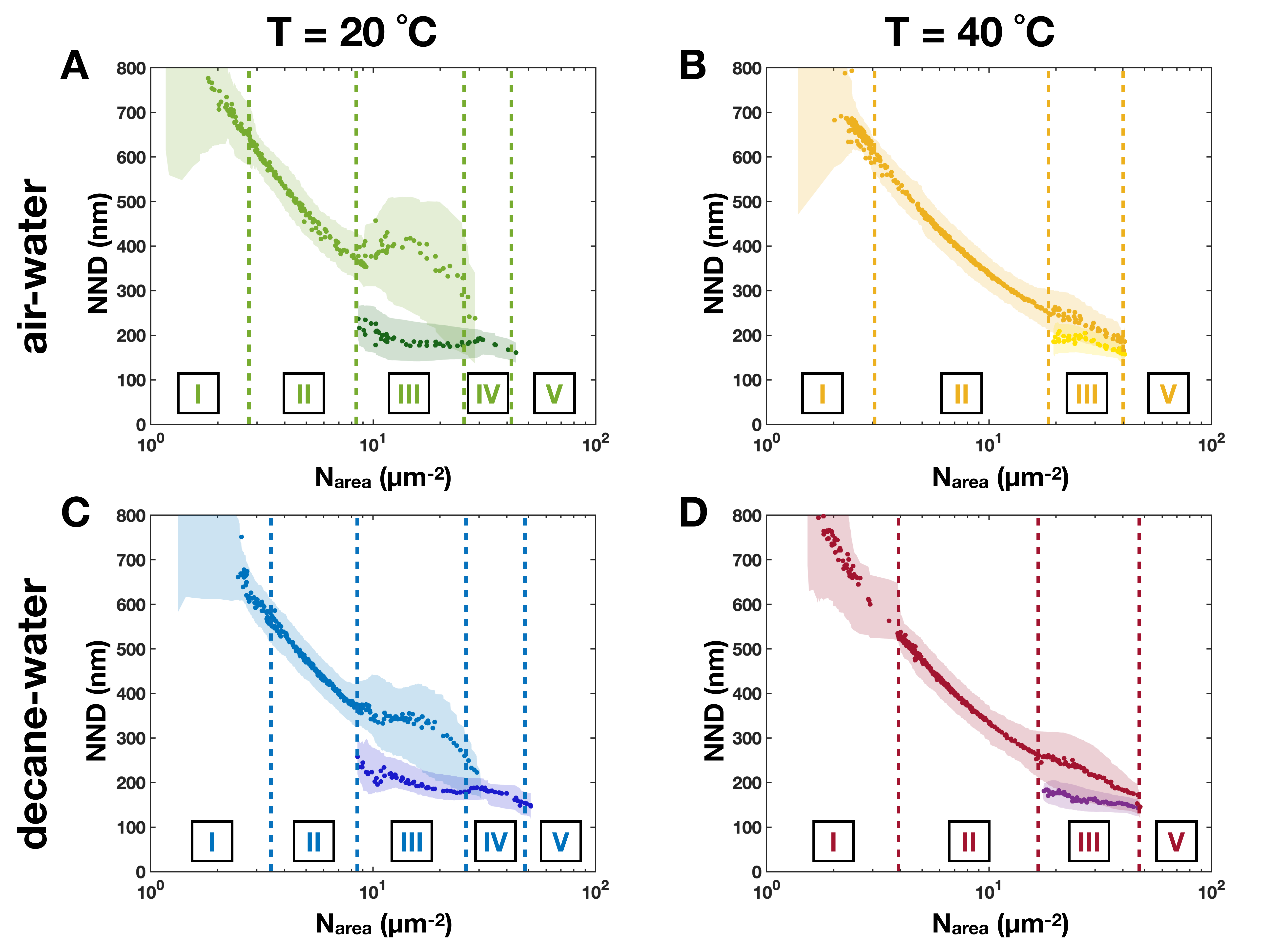}}
\centering
\caption{Mean nearest neighbor distances, \emph{NND}, versus number of microgels per area, $N_{area}$, of pNIPAM microgels at the air-water and decane-water interface. (A) T = 20~$^\circ$C, air-water interface. (B) T = 40~$^\circ$C, air-water interface. (C) T = 20~$^\circ$C, decane-water interface. (D) T = 40~$^\circ$C, decane-water interface. Shaded areas in the respective color represent the errors. Dashed lines indicate the transitions between different regimes. The regimes are labeled in roman numbers.
}
\label{fig:fig_NND}
\end{figure}

We follow the evolution of the in-plane diameter of the microgels within the monolayers at both interfaces more closely by plotting \emph{NND} as a function of $N_{area}$ (Figure~\ref{fig:fig_NND}). 
Starting from regime II a confluent monolayer is observed. The lateral compression with the barriers leads to a compression of the microgel coronae until in regime III the isostructural phase transition takes place (Figure~\ref{fig:fig_NND}A and B). Here, microgels in two phases coexist and the number of microgels in the second phase constantly increases with compression. Below the VPTT, regime IV is reached when all microgels have passed to the second phase. This layer can be compressed further. The monolayers above the VPTT do not display the regime IV and a direct transition from the third to the fifth regime is observed where the monolayers fail. \emph{NND}-values at characteristic interfacial concentrations are summarized in Table~\ref{tab:tab_NND}. 

Although the transition from regime I to regime II takes places at larger center-to-center distances, the isostructural phase transitions take place at nearly the same \emph{NND}, which is in good agreement with the size measurements of the single microgels (Fig.~\ref{fig:fig_AFM_SingleMicrogels}). For monolayers below the VPTT, the \emph{NND} at the transition from regime II to III is between 360-380~nm and above the VPTT between 250-270~nm. A different \emph{NND} for the air- and decane-water interfaces is also observed in regime V, when the monolayer is fully compressed and fails. Independent of temperature, the failure at the air-water interface is observed at a \emph{NND} of 160~nm, in contrast to 145~nm at the decane-water interface. This indicates that the amount of confined polymer per microgel increases with increasing surface tension of the interface. 

\begin{figure}[h!]
{\includegraphics[width=\textwidth]{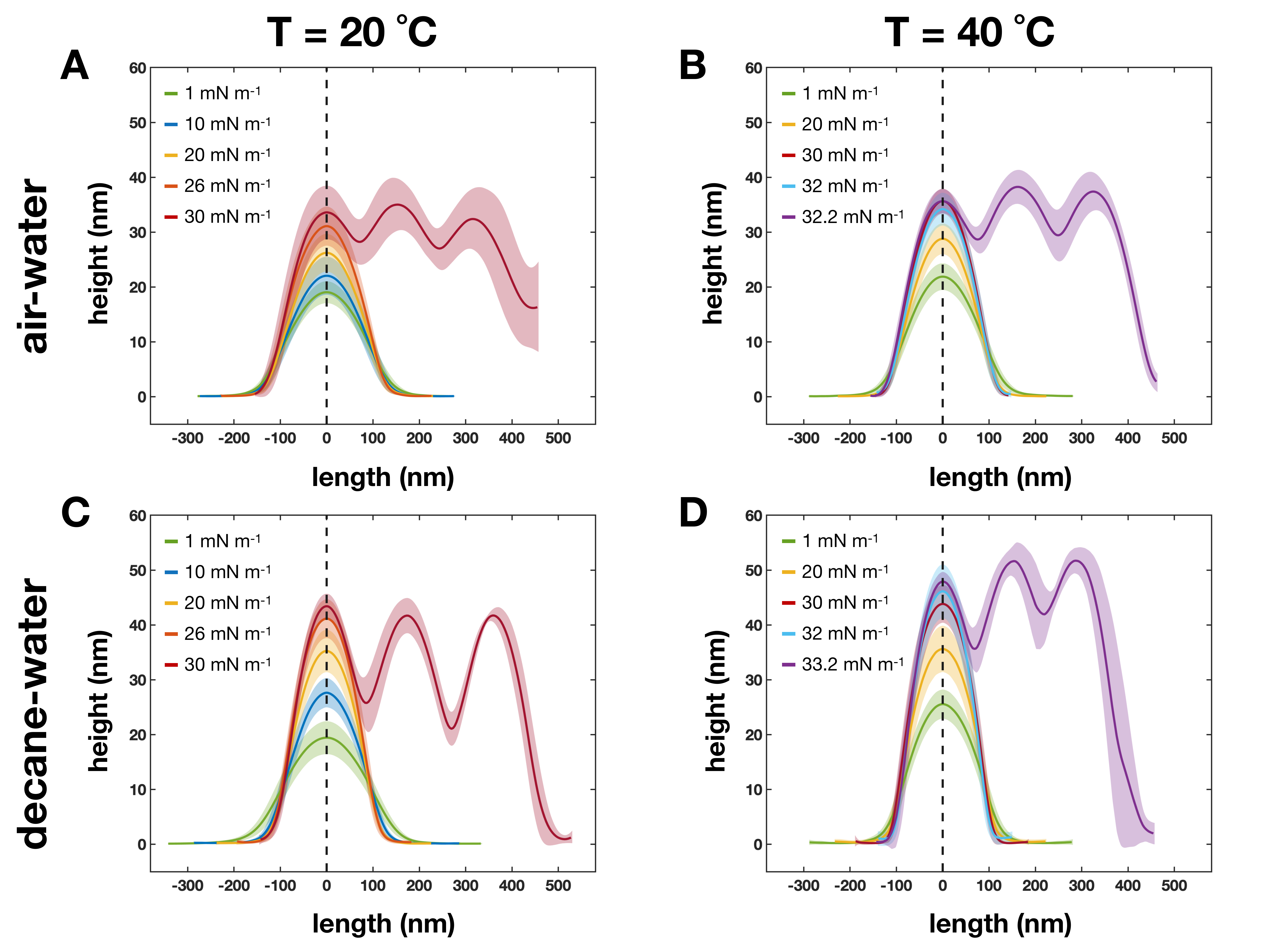}}
\centering
\caption{Averaged height profiles of microgels in dried state after deposition to a solid substrate at different temperatures, interfaces, and compressions. (A) T = 20~$^\circ$C, air-water interface. (B) T = 40~$^\circ$C, air-water interface. (C) T = 20~$^\circ$C, decane-water interface. (D) T = 40~$^\circ$C, decane-water interface.  Different colors represent different compressions. The standard deviation in height and length is given as shaded areas in the corresponding color. For each profile at least 40 microgels were analyzed.
}
\label{fig:fig_Height_Profiles}
\end{figure}

We also extract height profiles of microgels within monolayers to obtain a more detailed picture of their conformation under compression. The profiles of microgels at different compressions deposited from the air-water and oil-water and below and above the VPTT are presented in Figure~\ref{fig:fig_Height_Profiles}A-D. At the highest surface pressures, the microgel cores overlap and multiple microgels are shown. Here, $D_{core}$ was measured from the distance between the apices.  
The resulting values for $D_{core}$ and $H_{core}$ are summarized in Table~\ref{tab:tab_NND}. 

For all compressions, $H_{core}$ decreases from the decane-water to the air-water interfaces at the same temperature and $H_{core}$ increases from temperatures below to temperatures above the VPTT at the same top-phase. The later is related to the de-swelling of the microgel portions situated in the aqueous phase, although counter-intuitive. At the liquid interface, the portions situated in the aqueous phase still undergo a volume phase transition. At temperatures above the VPTT, they are de-swollen and display a more compact conformation with a smaller lateral extension and thickness. Since the amount of polymer within a microgel does not change during drying, the resulting profiles are flatter and more laterally extended below the VPTT, and higher and less laterally extended above the VPTT. 
 
The decrease in $H_{core}$ from the decane-water to the air-water interface suggests a more compact conformation of the microgel cores with a higher polymer density at and near the air-water interface. The height profiles only display the \emph{ex situ} structures and to fully understand the conformational changes experiments which resolve the \emph{in situ} structure are needed. Nevertheless, in addition with the aforementioned larger \emph{NND} in regime IV (Fig.~\ref{fig:fig_NND} and Tab.~\ref{tab:tab_NND}), the smaller $H_{core}$ at air-water interfaces gives a strong indication that the larger surface tension is compensated by adsorption of more polymer segments from the core of the microgels. 

Lastly, the ordering of the interfaces shows no substantial difference and the trend of $\Psi_6$ at the air-water interface (Figure S2A-B) and the decane-water interface (Figure~S2C-D) with compression is the same. Based on $\Psi_6$ and the characterization presented above, there is no evidence that a change in the top-phase influences the short- and long-range order of the pNIPAM microgel monolayers. A more detailed discussion of the evolution of both the hexagonal order parameter and the pair correlation functions, $G(r)$, with compression can be found elsewhere. \cite{Rey16, Boc19} 

\section{Conclusion}
In this study, we compared of pNIPAM microgels at two different interfaces: air-water and decane-water. Although, the two-dimensional phase behavior of microgels at air-water or decane-water is well known, so far, only qualitative comparison with the existing literature can be made. Here, we present a systematic study of the impact of the top-phase 
and elucidate quantitative differences between the confined microgels. The investigation is completed by measurements of the microgels, not only below but also above their VPTT. 

We show that the polymer network of microgels experience a different stretching at the two interfaces. Because both air and decane are non-solvents for pNIPAM, we attribute changes in their contact area to the differences in surface tension. The total diameter ($D_{2D}$) of the microgels increases from the decane- to the air-water interfaces, but the size of the microgel cores ($D_{core}$) is constant. We suggest that this is a consequence of the core-corona structure of microgels at the interface and associated with one or with a combination of three explanations:  
(i) in the center of a microgel the polymer network gives more resistance to deformation due to the higher cross-linker content, and experiences less stress due to the higher polymer content; 
(ii) Attenuation of the stretching for the microgels' fractions that are in the aqueous phase due to the flexibility of the polymer network; and (iii) Adsorption of more of the polymer network situated in the aqueous, \emph{i.e.}, out-of the interfacial plane, instead of stretching the polymer network in-plane.

We obtain strong evidences that more polymer of an individual microgels is adsorbed to the air-water interface considering the height profiles (Fig.~\ref{fig:fig_Height_Profiles}) and the inter-particle distance (\emph{NND}) close to the microgel monolayer failure (Fig.~\ref{fig:fig_NND}, regime IV). For the decane-water interface, it has been reported that the failure of the monolayer is independent of the temperature and that there is a maximum number of polymer segments that can be confined at or in close proximity of the interface. \cite{Boc19} When comparing the decane- and air-water interface, this limit is reached at larger \emph{NND} for the air-water interface. Additionally, height profiles of dried, deposited microgels are flatter at air-water interfaces than at oil-water interfaces; suggesting that the increase of interfacial tension is compensated by adsorption of more polymer segments of a single microgel to the interface. 

Furthermore, we use the obtained size of the individual microgels to estimate their elastic modulus between oil-water and air-water interfaces. In comparison to nanoindentation experiments normal to the interface, we determine a significantly larger modulus of the microgels in-plane ($\approx$ 23~MPa). This increase is attributed to both, the reduced amount of solvent in the polymer network close to the interface and the restrain of the polymer segments at the interface, in agreement with the literature. \cite{Fab20}

$\Pi$-$N_{area}$ Isotherms at the air-water interfaces at temperatures below and above the VPTT of the microgels show the same temperature-dependent changes as at the decane-water interface. At both interfaces, the temperature-dependent swelling perpendicular to the interface affects the compressibility parallel to the interface resulting in an isotherm with a single increase. Compared to the impact of temperature, the effect of the top phase (at the same temperatures) is negligible. 

In conclusion, our results unambiguously demonstrate that the two-dimensional phase behavior of microgels, in terms of their microstructure and their ordering, is not affected by having oil or air as top-phase. In contrast to a recent study at benzene- and hexane-water interfaces, \cite{Fab19} a dominant impact of the top-phase is observed for the in-plane stretching of the microgels corona, while their more cross-linked core and the microgels' portions situated in the aqueous phase are unaffected. 

\section{Acknowledgement}
The authors acknowledge financial support from the SFB 985 "Functional Microgels and Microgel Systems" of Deutsche Forschungsgemeinschaft within projects A3 and B8. This work is based upon experiments performed at the KWS-2 instrument operated by J\"ulich Centre for Neutron Science (JCNS) at the Heinz Maier-Leibnitz Zentrum (MLZ), Garching, Germany. We thank Marie Friederike Schulte and Maximilian Schmidt for fruitful discussions.

\end{singlespacing}
\bibliography{main_bib}

\end{document}